# Quantifying Epitaxial Growth using a Purely Topographical Signal

Kai Trepka, *UCSF School of Medicine, 533 Parnassus Ave, San Francisco, CA 94143*

*Abstract*—Thin films are ubiquitous, with uses ranging from optoelectronics to antibacterial coatings. Unfortunately, precisely quantifying how the choice of substrate influences epitaxial growth remains an unsolved problem. Here, a novel thin film of holmium oxide with record-high paramagnetic saturation was grown on a variety of substrates. Conventional attempts to extract epitaxial information to characterize the growth mechanism were ineffective, due to the unique size regime of the product. Instead, a signal-processing inspired Fourier method was used to elucidate information on epitaxial ordering from purely topographical data, avoiding the pitfalls of atomic-level diffraction. Further, we define and utilize an inner product-based metric termed a *q-score* that can quantify the relative degree of ordering of epitaxial crystallites. The q-score provides a direct measure of epitaxy, enabling more quantitative future studies of thin film growth.

## I. INTRODUCTION

Magnetic resonance force microscopy (MRFM) is an emerging technology that offers the promise of single nucleon detection in individual biological samples or nanodevices [1]. Such a device functions like a nanoscale magnetic resonance imaging (MRI) machine, applying a radiofrequency (RF) current and measuring response with a magnetic cantilever, all against the background of a strong uniform magnetic field [2,3]. Effective MRFM demands a large signal to noise ratio (SNR). From [3], this is:

$$SNR = \frac{N\mu_N^2}{2\Delta f k_B T}\left[\left(\frac{\partial B}{\partial x}\right)^2 \left(\frac{w_c Q}{k_c}\right)\right] \quad (1)$$

Since the SNR is quadratically dependent on the magnetic field gradient $\frac{\partial B}{\partial x}$, developing strong, new magnetic materials is important for optimizing MRFM. Recently, a new crystal phase of holmium oxide thin film with record-high paramagnetic saturation (above 2 Tesla) was synthesized using a thermal physical vapor deposition technique [4]. Holmium films grown on different substrates (A-plane sapphire, C-plane sapphire, and amorphous quartz) under otherwise identical deposition conditions have different morphological properties, suggesting that the substrate controls film growth. This direction is not a chemical process, as the two sapphire substrates are chemically identical, and growth occurs in a regime well below the vaporization of all the substrates [5,6]. Rather, the different crystal structures and orientations of the substrates visually appear to direct holmium growth in different ways. For example, in Figure 1, we see the holmium oxide film grows into crystalline triangles regardless of surface, but on the Sapphire C substrate (Fig. 1b), they seem to be aligned with one another.

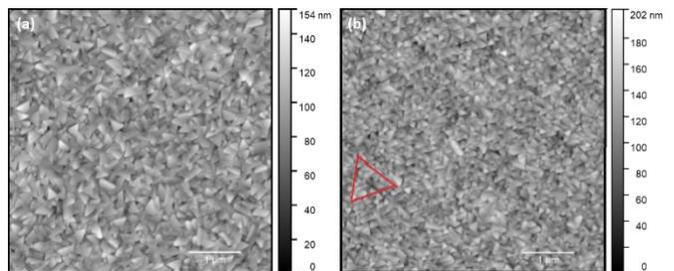

*Fig. 1. Atomic force micrographs of holmium oxide thin film growth on Sapphire A (a) and Sapphire C (b) substrates. Upon visual inspection, the triangular crystallites appear to be randomly oriented with respect to one another in (a), but roughly aligned with the drawn red triangle in (b).*

When thin films grow, individual atoms land on the surface and then move around until they lose their kinetic energy and reach a thermodynamic minimum [7]. This can occur in different ways depending on how the landing atoms align with the substrate crystal atoms. When films grow on a substrate, the substrate can direct their growth to align with its own crystal structure, creating a periodic result [8]. This is termed *epitaxial growth*.

To determine whether different substrates are directing growth of different crystal phases of holmium oxide, we attempted a variety of standard diffraction techniques, including coarse x-ray diffraction (XRD) and more fine-grained elastic recoil detection (ERDA) and transmission electron microscopy (TEM). Most of these techniques determine bulk crystal structure and elemental composition, ultimately determining that we have a new phase of holmium oxide. However, they operate at the wrong size scale to determine whether the crystallites themselves are oriented. XRD and ERDA are too coarse to detect individual crystal grains reliably [9,10]. Available TEM only operates at a small domain size (~50 nm), while the crystallites in question are on the order of 100 nm. Although TEM has been successful at studying epitaxial growth in the past, it has traditionally been done on much thinner and smaller nanostructures [11] or been focused on defects [12]. In other words, no techniques were successful at revealing the orientation of individual crystal grains relative to *each other*, which is key to understanding the influence of using different substrates on holmium oxide growth and designing future experiments.

Fortunately, techniques from signal processing can be used to uncover periodicity that is otherwise not apparent in images. In particular, the 2-dimensional Fourier transform is a promising tool and is defined by the relation in Equation (2).

While the most common transforms in signal processing work by converting between time and frequency domains, the 2D transform here operates between two spatial regimes: real, physical space, parametrized by $x, y$, and inverse *k-space*, parametrized by $u, v$. By converting the image from real space into k-space using an orthogonal basis of complex exponentials, patterns emerge that can be used to quantify how ordered the substrate is [13]. In particular, the discrete Fourier transform (DFT, defined by Equation (3)) of an image is readily implemented using an FFT algorithm in MATLAB, as described in [14].

$$F(u,v) = \int_{-\infty}^{\infty}\int_{-\infty}^{\infty} f(x,y) e^{-j2\pi(xu+yv)}\, dxdy \quad (2)$$

$$F[u,v] = \frac{1}{XY} \sum_{x=0}^{X-1}\sum_{y=0}^{Y-1} f[x,y] e^{-j2\pi\left(\frac{u}{X}x+\frac{v}{Y}y\right)} \quad (3)$$

Note that in (3), X is the total number of pixels in the horizontal direction, and likewise for Y in the vertical direction. The discrete transform sums over all available pixels and sets $f[x,y] = 0$ outside the observed X by Y region.

Quantifying epitaxy has applications beyond holmium oxide thin films. Thins films are ubiquitous [15], with applications ranging from optoelectronics as in [16] to antibacterial coatings as in [17]. In general, quantifying the influence of substrate on thin film growth, termed *epitaxial growth,* is a challenging problem [8,18]. Most literature relies on expensive atom-level measurements and can only detect a few layers of atoms (~1 nm) worth of epitaxial growth [11,12,19], while thin films with practical applications are often over 100 nanometers thick [20].

Here, we develop and test a facile classification scheme that quantifies how ordered thin film growth is using only topological data. This approach relies on transforming topographical atomic force microscopy (AFM) data into k-space and comparing the result to the predicted Fourier transform of perfect thin film epitaxy, as well as measuring maintenance of periodicity over large regions of film.

The following notation will be used throughout the manuscript:
- $f[x, y]$ denotes the z-height/brightness at a given spatial coordinate of an AFM image
- $F[u, v]$ denotes the transform of $f(x, y)$; this is the FFT in all cases except in section V.
- $< A, B >$ denotes a Frobenius inner product
- $A^*$ denotes the complex conjugate of $A$

## II. TOPOGRAPHICAL CRYSTALLINITY

It is well known that using different substrates can result in the growth of different types of crystal grains. There are three main cases of thin film growth, reviewed extensively in [8] and [15].
1. *Epitaxial crystalline growth*. In the case of epitaxial growth, the crystal grains align perfectly with the substrate's crystal structure, and the edges of crystal grains also align with the substrate and thus align with each other. Topographical data of this type of growth presents as a perfect tessellation of crystal edges. The shape of the grains depends on the shape of the substrate's exposed crystal plane.
2. *Non-epitaxial crystalline growth*. In the case of non-substrate directed crystalline growth, nucleation and growth occurs in random directions. Although crystals with hard edges can still form, they will not be aligned with each other. This can be modeled as an assortment of randomly oriented crystals, where the shape of the crystal depends mostly on the thin film material's intrinsic crystalline properties.
3. *Random growth*. In the case of non-epitaxial, non-crystalline growth, the thin film grows haphazardly on an unordered substrate, forming unordered structures. This case is largely non-instructive, and has no consistent morphological ordering.

In this experiment, holmium oxide was grown on two different crystal planes of sapphire, A-plane and C-plane, depicted in Fig. 2. The A-plane atomic cross section is a rectangular lattice, while the C-plane cross-section contains a hexagonal lattice. Meanwhile, holmium naturally grows into a hexagonal lattice [21] that is a close match in terms of lattice strain to C-cut sapphire, so it is expected that the C-plane sapphire will clearly direct its growth, whereas holmium will grow into randomly oriented triangular or hexagonal crystallites on the A-plane sapphire. The amorphous quartz substrate is not crystalline, and thus cannot epitaxially direct growth.

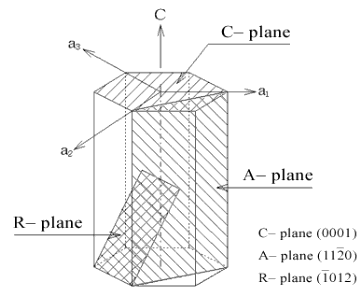

*Fig. 2. Different sapphire crystal planes. C-plane sapphire is a hexagonally symmetric lattice, while A-plane sapphire is more rectangular.*

Since there should be one main set of periodic components in the first case, and many rotated sets of the same kind of triangle in the second case, the 2-dimensional Fourier transform (FT) offers a way to extract the magnitude of different periodic components and quantify underlying periodicity. The 2D FT is based on the same principles as a one-dimensional Fourier transform, using complex exponential functions of x and y as the basis set.

Since we have discrete images of 512x512 datapoints, we have to use the DFT described in Equation (3) rather than the FT described in Equation (2), with $X = Y = 512$. Note that this transform works theoretically because the AFM signal is viewed as a multiplication of the real thin film surface with the *bed-of-nails* function in Equation (4), which samples out 262,144 discrete datapoints from the true, underlying, continuous thin film surface.

$$\sum_{n=-\infty}^{\infty}\sum_{m=-\infty}^{\infty}\delta(x-nX)\delta(y-mY) \quad (4)$$

Implementing the FFT in MATLAB on a representative set of potential crystal patterns, it is apparent that the FFT of disordered triangles contains the superposition of many rotated transforms of individual triangles, ultimately creating a messy (and, *ad infinitum*, uniform and circularly symmetric) Fourier transform (Fig. 3a-b).

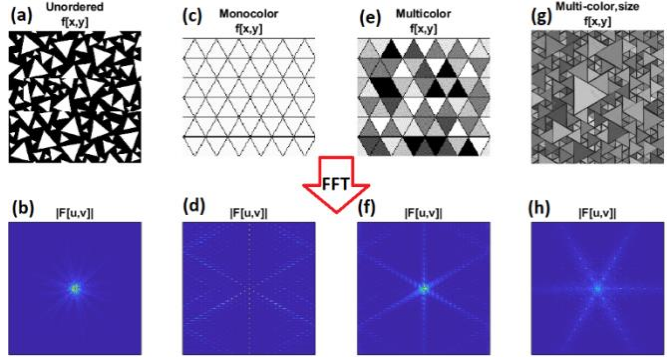

*Fig. 3. Theoretical Fourier domain images of ordered (epitaxial) vs disordered (non-epitaxial) triangles. (a), (c), (e), and (g) depict initial images, while (b), (d), (f), and (h) are their respective FFTs. In (a) and (b), the unordered triangles create a mess of overlapping, rotated hexagons. In (c) and (d), a uniform, perfectly ordered tessellation of identical equilateral triangles creates rhomboidal patterns out of discrete dots. In (e)-(h), tessellations of triangles of varying height and size both create hexagonal FT patterns.*

Meanwhile, any set of ordered or tessellated triangles results in six main lines originating from the zero-order center, rotated depending on the initial orientation of the triangles (Fig. 3c-h). The hexagonal symmetry emerging from a central node is most apparent when multicolored triangles are used (Fig. 3e-h). These multicolored images are more realistic representations of crystal grain growth, individual grains grow to varying sizes and thus are different heights in AFM micrographs.

The hexagonal ordering and Fourier pattern appear to be independent of variations in amplitude (color) or size variations in the initial triangle tessellation, suggesting the Fourier transform may be an effective way to classify ordering in crystal growth. Now that it is clear that the different theoretical growth regimes of crystals should be distinguishable using the Fourier transform, the given holmium oxide AFM data can be transformed into k-space and qualitatively compared to the theoretical transforms in Fig. 3.

### III. QUALITATIVELY INVESTIGATING HOLMIUM OXIDE THIN FILM GROWTH

Taking the 2-dimensional FFT of the holmium oxide thin film data yields k-space images that immediately reveal orientation differences between the films grown on A-plane sapphire, C-plane sapphire, and amorphous quartz.

In particular, both films grown on sapphire substrates have considerable spread from the zero-order lines in the Fourier plane (Fig. 4a-d), while the film grown on quartz is clearly disordered optically and has no visible triangles (Fig 4e). The Fourier transform of the film grown on amorphous quartz is focused around the zero-order lines, suggesting more random growth (Fig. 4f).

Note that in the initial AFM image, while it is clear to the eye that the crystallites are triangular on both A sapphire (Fig. 4a) and C sapphire (Fig. 4c), whether these triangles are aligned with each other is an open question. However, the slight qualitative differences in the Fourier domain suggest an answer. While the transform of the holmium oxide film grown on Sapphire A is largely circularly symmetric (Fig. 4b), suggesting some amount of sharp triangular edges but not ordered triangles (similar to the pattern in Fig. 3b), the transform of the holmium oxide film grown on C-plane sapphire has higher brightness along a hexagonal set of contours (Fig. 4d), akin to the theoretical transforms in Fig. 3f and 3h.

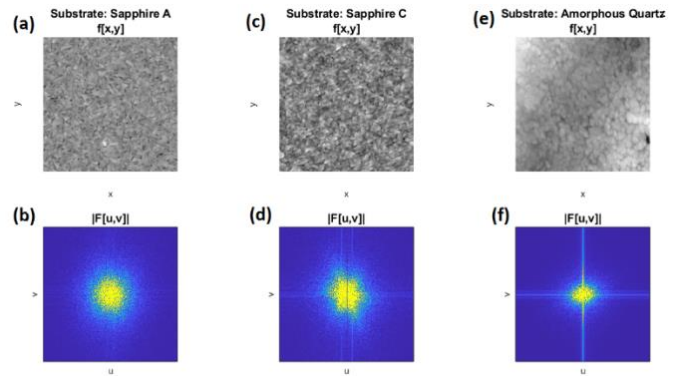

*Fig. 4. Holmium oxide thin films grown on three different substrates under identical run conditions and their respective FFTs. The thin film on sapphire A (a), which contains triangular crystallites, has no clear ordering in its FFT (b). However, the thin film on sapphire C (c), which also contains triangular crystallites, has hexagonal ordering in its FFT (d), suggesting crystallite alignment. To contrast, there is no clear ordering in either the AFM image (e) or FFT (f) of the film grown on amorphous quartz, only large zero-order patterns and symmetric spread.*

While this qualitative comparison of Fourier transforms to the theoretical transforms of tessellations given in Fig. 3 suggests a slight difference in epitaxial ordering between the Sapphire A- and Sapphire C-grown films, it is unclear if there is some fine degree of overall threefold ordering on Sapphire A (which would be possible if, for example, each triangular crystallite constrained the orientation of the neighboring crystallite; in other words, the holmium could self-direct its own growth.) Thus, quantifying the perceived difference between Fig. 4b and Fig. 4d is important for understanding how ordered the thin films are.

### IV. QUANTIFYING EPITAXIAL GROWTH

At present, two tools are widely used quantify the influence of epitaxy on nanostructure and thin film growth.
1. Direct measurements using TEM, as in [11] and [12]. This reveals the degree of lattice mismatch at the substrate-film interface and is thus the ideal tool for quantifying epitaxial growth. However, this tool is ineffective when the resulting thin film does not perfectly match up with the sapphire crystal lines (as with the holmium oxide), and when the size regime of interest extends beyond a couple nanometers (in our case, the relevant regime is hundreds of nanometers; see Fig. 1).

2. Theoretical lattice mismatch [22]. Every material has a known crystal structure, and the difference between lattices (i.e. the size of a holmium hexagon vs a sapphire hexagon) can be computed. Lattice strain $s$ is determined by the crystal spacing of the substrate $a_{sub}$ and intrinsic crystal spacing of the film material $a_{epi}$ in Equation (5). Typically, a lattice strain of below 10% ($s = 0.1$) suggests epitaxial thin film growth.

$$s = \frac{a_{sub} - a_{epi}}{a_{sub}} \quad (5)$$

However, this technique is not a true "measurement" of ordering, but rather a convenient percentage that can be reported along with qualitative images of clearly ordered growth (i.e. aligned nanowires). Lattice mismatch does not always determine epitaxial growth and vice versa [23]. For example, the lattice strain between holmium oxide and sapphire C is $s = 0.25 > 0.1$, but it is clear by inspection that choice of substrate influences growth (Fig. 4a,c,e).

As a result, development of a quantification scheme beyond lattice strain that can utilize easy-to-collect topographical data (i.e. from AFM, SEM) is paramount to understanding partial ordering and ordering on large scales.

Once again, tools from signal processing come into play. Rather than qualitatively comparing the FFT of an ordered lattice of triangles with that of AFM images, we can quantitatively compare the two using an inner product.

First, we choose a *quantifier*, an ideal image and corresponding FFT to model the epitaxy that we would like to measure a given thin film's similarity to. Note that the choice of quantifier will define our result, so future work must be careful when comparing different results to make sure the same choice of quantifier image was used.

To measure the similarity of a given image with the quantifier, we could simply take a Frobenius inner product (Equation (6)) of the matrix $Q$ representing the FFT of the quantifier and $A$ representing the FFT of the image. If the hexagonal patterns align, this will be larger than if none of the patterns align.

$$< A, Q > = \sum_{i,j} A^*_{ij} Q_{ij} \quad (6)$$

However, in order for this computation to work, the FFTs must be pre-processed first. Otherwise, images that are very bright will automatically have a very large inner product, or triangles that are not exactly aligned with those of the quantifier image may have a low inner product despite perfect alignment between triangles in the AFM image. The *quantifier algorithm* is as follows:

1. Load the images and take their FFTs.
2. Remove uniform background noise.
3. Remove outliers above an experimentally determined high quantile. In the future, an LPF could be used.
4. Normalize each FFT $F$ by dividing by its Frobenius norm, $\sqrt{<F,F>}$, where the inner product is given in (6).
5. Compute what we here define as the *q-score*, defined as the inner product of the processed FFTs $A', Q'$:

$$q = < A', Q' > \quad (7)$$

6. In addition, compute the error, defined below. Note that low q-scores are correlated with high errors.

$$\epsilon = \sum_{i,j} |A_{ij} - Q_{ij}|^2 \quad (8)$$

There is one further challenge: rotating an image will also rotate the FT of that image. For instance, rotation of the quantifier triangle tessellation image by 10 degrees will also rotate its Fourier transform by 10 degrees. The proof is below:

*Working in polar coordinates, set $x = r\cos\theta, y = r\sin\theta, u = \rho\cos\phi, v = \rho\sin\phi$. Plugging this into Eq. 2, we see $F(\rho, \phi)$*

$$= \int_0^\infty \int_0^{2\pi} f(r,\theta) e^{-j2\pi(r\cos\theta\,\rho\cos\phi + r\sin\theta\,\rho\sin\phi)} \, r\,dr\,d\theta$$

$$= \int_0^\infty \int_0^{2\pi} f(r,\theta) e^{-j2\pi r\rho\cos(\theta-\phi)} \, r\,dr\,d\theta$$

*By inspection, a rotation in real space results in the same rotation in k-space, i.e.*

$$f(r, \theta + \theta_0) \leftrightarrow F(\rho, \phi + \theta_0) \quad (9)$$

In order to avoid accidental mismatch between potential perfect tessellations of the substrate and quantifier (i.e. the substrate hexagonal pattern could, by chance, be rotated 30 degrees away from the quantifier FFT hexagonal pattern), we will rotate the quantifier by small increments, going through the above algorithm and computing the q-score and error for each rotation. We can perform this rotation in real space and then take the Fourier transform, as rotating in either basis is equivalent by (9). Then, we select the highest q-score (i.e. the best match) and set that as the true q-score, along with its concurrent error. Note that for a 6-fold symmetric image such as a tessellation of triangles, we only have to rotate through a total of 60 degrees before the original pattern returns, decreasing computational time. In addition, the amount of each rotation can be tuned according to computational demands.

A quick internal calibration of the quantification algorithm is that it should return 1 if $Q = A$; this is indeed the case.

Next, we test this algorithm on our images, depicted in Figure 5. The different quantifiers used are the mono-color image (Fig. 1c), multi-color, -size image (Fig. 1g), and multicolor image (Fig. 1e). Recall that a high q-score is indicative of high matching with the ideal pattern, and subsequent high ordering. While the particular q-scores change depending on the choice of quantifier, the ordering remains the same: sapphire C is the most ordered, followed by sapphire A. Amorphous quartz is the least ordered.

In addition, note that the multicolor, multi-size triangle tessellation quantifier offers the greatest distinguishability. This is expected, given that it is most similar to the AFM pattern that would actually be observed in the case of epitaxial growth: crystallites are all oriented, but can vary in height and lateral size, represented by variations in color and size respectively.

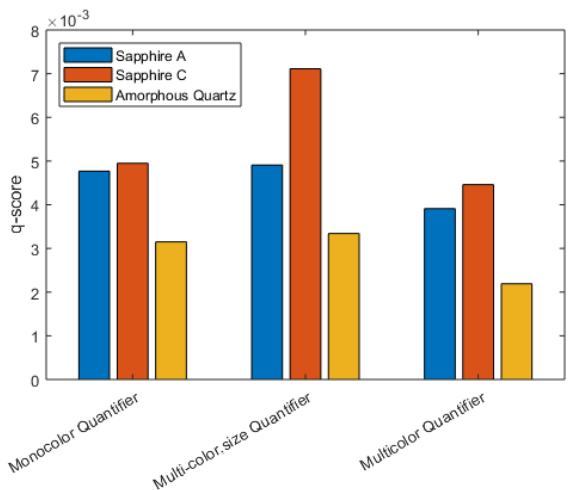

*Fig. 5. Ordering of thin-film epitaxy on various substrates using the q-scores. Different quantifiers are used to compute different q-scores, but all 3 give the same relative ordering, suggesting that sapphire C is indeed more epitaxially ordered than sapphire A, and both growths on sapphire are more ordered than the growth on amorphous quartz.*

One additional use of this quantification scheme lies in comparing images to one another. For example, one other set of data we have is images from each substrate taken 5 mm apart from one another, a measure of long-range ordering. We can then set each image as Q, and define another image taken from the same substrate as A; in other words, we are taking a version of the inner product between images from different portions of a given substrate. If there is long-range ordering, this q-score should be large; otherwise, it will be small.

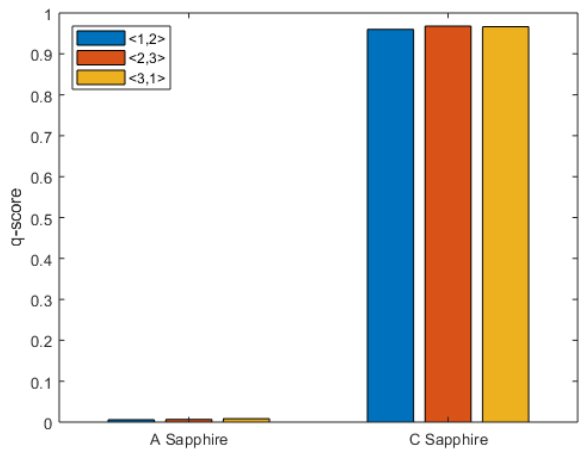

*Fig. 6. q-scores of long-range ordering on different substrates. A higher q suggests C has more long-range ordering than A.*

Using 3 images from each of sapphire A and sapphire C, each image taken 5 mm from the last along the same line of translation, we compute the q-scores between each permutation of the 3 images. Zero rotation of FFTs is used, as rotation occurring over the course of translation would indicate that the long-range ordering is broken. The resulting q-score for sapphire C is much higher than that of sapphire A (Fig. 6). Thus, the film grown on sapphire C is well-aligned across enormous spatial jumps (5 mm >> 5 microns, the span of each image). This preservation of ordering over a long range is highly unlikely to occur unless the underlying substrate is directing growth at each point, making this a direct and quantitative measure of epitaxy.

The error terms, computed using Equation (8), are small for sapphire C, but large for sapphire A (Fig. 7), confirming a poor match in Fourier space between different images from sapphire A and consequent non-epitaxial growth.

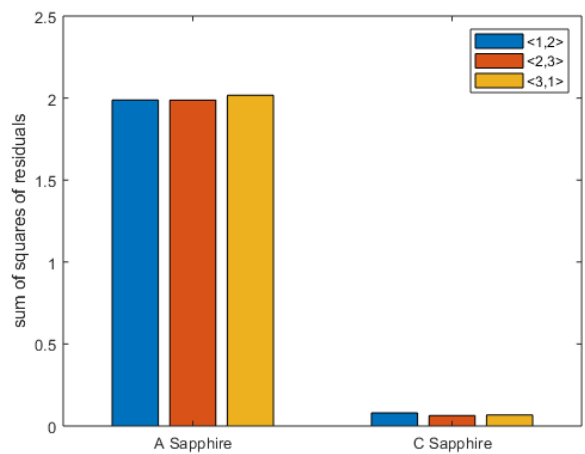

*Fig. 7. Computed error of matching for long-range ordering. Sapphire A has much higher error, suggesting less order and less epitaxy.*

One interesting result is that the q-score, when comparing shifted images, approaches 1 for the aligned growths on sapphire C (Fig. 6). This approach utilizes the shifted AFM images as quantifiers. However, the q-score is nearly 1000 times smaller when using an abstracted perfect tessellation as the quantifier (Fig. 5). As a result, the best use of the quantifier algorithm is likely to compute *self* q-scores, i.e. quantify the difference in Fourier transforms between different regions of a single thin film to detect long-range ordering. This use alleviates one of the biggest problems with the quantifier algorithm: the choice of the ideal quantifier or FFT to compare our signal to. From Figure 5, it is apparent that altering quantifiers can change q-scores by nearly 50%, and that differentiability is sometimes a challenge (i.e. the q-scores of sapphire A and sapphire C are very close for all but the multicolor, multi-size quantifier). Comparing images to their downstream counterparts adds an internal control, which is desirable as the exact pattern of epitaxial growth is unknown due to uncertainties at nanoscale.

## V. THE DISCRETE COSINE TRANSFORM: ANOTHER WEAPON IN THE ARSENAL?

The success of using the q-score to quantify long-range ordering relies on the transformation of real space to a form of frequency space. Although the Fourier transform was used in the quantifier algorithm, alternative transforms could potentially be used. For example, the discrete cosine transform (DCT) is often used for lossy image compression, relying on an orthogonal basis of cosine functions (rather than complex exponential functions, as with the FT) [24]. The DCT is defined below, omitting normalizing factors:

$$F(u,v) \propto \sum_{x=0}^{X-1} \sum_{y=0}^{Y-1} f(x,y) \cos\left[\frac{(2x+1)u\pi}{2X}\right] \cos\left[\frac{(2y+1)v\pi}{2Y}\right] \quad (9)$$

We define our *DCT quantifier algorithm (DQA)* as identical to the previous algorithm, except taking a 2D DCT in all places where a 2D FFT was previously performed. Applying the DQA to a few test cases shows that it is potentially useful, but quantitatively challenging to optimize.

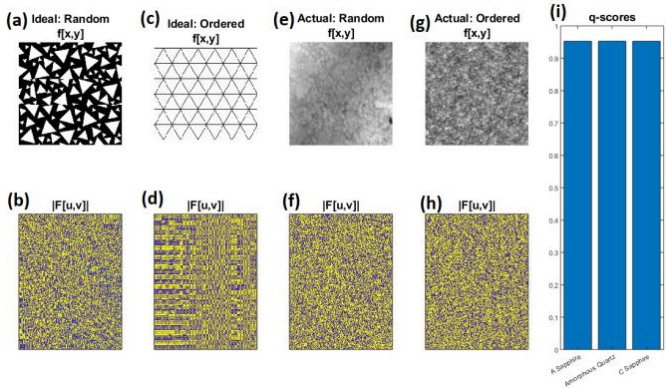

*Fig. 8. Applying DCT to epitaxial characterization. (b) depicts the DCT of (a), while (d) depicts the DCT of (c). The ideal ordered triangles have much more "striped" DCT than the randomly oriented triangles. However, this qualitative distinction disappears when comparing thin films grown on sapphire A (e,f) and sapphire C (g,h). (i) depicts the q-scores for the DQA on all substrates; they are virtually identical.*

When looking at the discrete cosine transforms of ideal ordered vs unordered triangles (Fig. 8a-d), differences are immediately apparent. The DCT of the unordered triangles in (8a) appears random (8b), while that of the triangular tessellation in (8c) appears to contain more slowly varying outputs and longer regions with the same coefficients, resulting in a striped pattern (8d).

This theoretical qualitatively useful difference disappears when looking at real samples. The transform of the totally random sample on quartz (8e) looks like a pattern of distorted static (8f) not so different than the transform of ordered holmium thin film on sapphire c (8h). Differences invisible to the eye often show up when computing computational similarities when using an inner product; unfortunately, applying the DQA to the three substrates results in virtually identical q-scores (8i), suggesting little quantitative difference when using cosines as the orthogonal basis set for transforming.

Implementation of the discrete wavelet transform (DWT, reviewed in [25]) produces similar results. There are differences between the ideal ordered and random cases, but little immediate quantitative difference with the real films.

One final future strategy beyond the scope of this work is development of a neural network for image recognition of FFTs that can distinguish between epitaxial, partially epitaxial, and non-epitaxial growth [26], which would improve upon the q-score as a metric. Although work has been done to optimize such a network in the case of particular individual thin film materials' growth conditions [27], a larger project could utilize the FFTs of a wide array of datasets known by TEM or other techniques to demonstrate epitaxy, partial epitaxy, or random growth to train the algorithm. This approach would also allow quantification of epitaxy from relatively low-resolution topographical images alone.

## VI. CONCLUSION

Because of the broad applications of epitaxially-grown thin films, mechanistically understanding their growth is critical to developing and characterizing new materials. In particular, a set of holmium oxide thin films grown on three different substrates under otherwise identical conditions were found to be exceptionally paramagnetic, but the degree of epitaxy was not measurable via traditional diffraction techniques, inhibiting mechanistic understanding of their growth. However, it was possible to collect extensive topographic data via AFM, and it was obvious that the holmium oxide crystallites appear to exhibit symmetry akin to that of a triangular tessellation.

Here, we developed a qualitative theoretical framework for transformations of ordered vs unordered triangles into k-space, demonstrating 6-fold symmetry for the ordered case and symmetric spreading for unordered triangles. We compared the FFTs of holmium oxide crystallites grown on various substrates to the transforms of these ideal cases, which suggested that growths on sapphire C are epitaxially aligned.

Next, we developed and tested an algorithm for calculating a "q-score", a modified inner product between FFTs that quantifies the amount of similarity (and thus epitaxial ordering), confirming that sapphire C was indeed more ordered than sapphire A. Since this q-score relies on comparing two images, two approaches can be used: comparing a theoretically ideal transform to any image or comparing different images from the same substrate. The former allows comparison and potential identification of triangular vs other types of symmetry, while the latter allows direct quantification of long-range epitaxy with a robust internal control. This algorithm was modified to operate using DCTs and DWTs, with some results but little practical quantification success.

Ultimately, the C-sapphire substrate is most effective at epitaxially directing holmium oxide thin film growth, making it a candidate for future experiments. More broadly, utilizing Fourier space enables quantification of epitaxial growth in the absence of crystal data. The q-score as a tool measures and quantifies epitaxial growth, allowing mechanistically guided development of new materials.

## VII. REFERENCES

[1] D. Rugar, R. Budakian, H. J. Mamin, and B. W. Chui. "Single spin detection by magnetic resonance force microscopy." *Nature*, vol. 430, no. 6997, pp. 329, 2004.

[2] J. Sidles, J. Garbini, K. Bruland, D. Rugar, O. Zuger, S. Hoen, and C. Yannoni. "Magnetic resonance force microscopy." *Reviews of Modern Physics*, vol. 67, no. 1, pp. 249, 1995.

[3] C. Degen, M. Poggio, H. Mamin, C. Rettner, and D. Rugar. "Nanoscale magnetic resonance imaging." *Proceedings of the National Academy of Sciences*, vol. 106, no. 5, pp. 1313-1317, 2009.

[4] K. Trepka, R. Hauert, C. Cancellieri, and Y. Tao. "A robust metal oxide thin film with saturation magnetization exceeding 2 Tesla." *Matter*, vol. 3, no. 4, pp. 1263-1274, 2020. doi:10.1016/j.matt.2020.07.031

[5] B. Cockayne, M. Chesswas, and D. Gasson. "Single-crystal growth of sapphire." *Journal of Materials Science,* vol. 2, no. 1, pp 7-11, 1967.

[6] N. Ainslie, J. Mackenzie, and D. Turnbull. "Melting kinetics of quartz and cristobalite." *The Journal of Physical Chemistry*, vol. 65, no. 10, pp. 1718-1724, 1961.

[7] Z. Zhang and M. Lagally. "Atomistic processes in the early stages of thin-film growth." *Science,* vol. 276, no. 5311, pp. 377-383, 1997.

[8] J. Narayan and B. C. Larson. "Domain epitaxy: A unified paradigm for thin film growth." *Journal of Applied Physics*, vol. 93, no. 1, pp. 278-285, 2003.

[9] A. Monshi, M. Foroughi, and M. Monshi. "Modified Scherrer equation to estimate more accurately nano-crystallite size using XRD." *World Journal of Nanoscience and Engineering*, vol. 2, no. 3, pp. 154-160, 2012.

[10] H. Angerer, D. Brunner, F. Freudenberg, O. Ambacher, M. Stutzmann, R. Höpler, T. Metzger, et al. "Determination of the Al mole fraction and the band gap bowing of epitaxial Al x Ga 1− x N films." *Applied Physics Letters*, vol. 71, no. 11, pp. 1504-1506, 1997.

[11] M. Hammar, F. K. LeGoues, J. Tersoff, M. C. Reuter, and R. M. Tromp. "In situ ultrahigh vacuum transmission electron microscopy studies of hetero-epitaxial growth I. Si (001) Ge." *Surface Science,* vol. 349, no. 2, pp. 129-144, 1996.

[12] T. Metzger, R. Höpler, E. Born, O. Ambacher, M. Stutzmann, R. Stömmer, M. Schuster et al. "Defect structure of epitaxial GaN films determined by transmission electron microscopy and triple-axis X-ray diffractometry." *Philosophical Magazine A*, vol. 77, no. 4, pp. 1013-1025, 1998.

[13] B. Osgood. "Lecture notes for EE 261: the Fourier transform and its applications." Electrical Engineering Department, Stanford University, 2013.

[14] W. Huang and D. MacFarlane. "Fast Fourier transform and MATLAB implementation." 2016. Available online at https://www.utdallas.edu/~dlm/3350%20comm%20sys/FFTandMatLabwanjun%20huang.pdf.

[15] M. Ohring. Materials science of thin films. *Elsevier*, 2001.

[16] R. Davis, G. Kelner, M. Shur, J.W. Palmour, and J.A. Edmond. "Thin film deposition and microelectronic and optoelectronic device fabrication and characterization in monocrystalline alpha and beta silicon carbide." *Proceedings of the IEEE,* vol. 79, no. 5, pp. 677-701, 1991.

[17] H.J. Jeon, S.C. Yi, and S.G. Oh. "Preparation and antibacterial effects of Ag–SiO2 thin films by sol–gel method." *Biomaterials,* vol. 24, no. 27, pp. 4921-4928, 2003.

[18] K. Fichthorn and M. Scheffler. "Island nucleation in thin-film epitaxy: a first-principles investigation." *Physical Review Letters*, vol. 84, no. 23, pp. 5371, 2000.

[19] S. Emori, D. Yi, S. Crossley, J.J. Wisser, P.P. Balakrishnan, B. Khodadadi, P. Shafer et al. "Ultralow Damping in Nanometer-Thick Epitaxial Spinel Ferrite Thin Films." *Nano Letters,* vol. 18, no. 7, pp. 4273-4278, 2018.

[20] C. Tellier and A. Tosser. Size effects in thin films. Vol. 2. *Elsevier*, 2016.

[21] A. Ghorai. "Calculation of Parameter of the Ashcroft Model Potential for Hexagonal Closed Pack (hcp) Crystals." *Acta Physica Polonica*, no. 2, pp A134, 2018.

[22] A. Trampert, and K. H. Ploog. "Heteroepitaxy of Large-Misfit Systems: Role of Coincidence Lattice." *Crystal Research and Technology: Journal of Experimental and Industrial Crystallography,* vol. 35, no. 6-7, pp. 793-806, 2000.

[23] Y. Chen and J. Washburn. "Structural transition in large-lattice-mismatch heteroepitaxy." *Physical Review Letters,* vol. 77, no. 19, pp. 4046, 1996.

[24] H. Ochoa-Dominguez and K. R. Rao. Discrete Cosine Transform. *CRC Press*, 2019.

[25] A. Jensen and A. Cour-Harbo. Ripples in mathematics: the discrete wavelet transform. *Springer Science & Business Media*, 2001.

[26] K. Simonyan and A. Zisserman. "Very deep convolutional networks for large-scale image recognition." *arXiv preprint* arXiv:1409.1556, 2014.

[27] F. Karimzadeh, A. Ebnonnasir, and A. Foroughi. "Artificial neural network modeling for evaluating of epitaxial growth of Ti6Al4V weldment." *Materials Science and Engineering: A*, vol. 432, no. 1-2, pp. 184-190, 2006.